\documentclass{raa_twocolumn}

\usepackage{natbib}
\usepackage{verbatim}
\usepackage{color}
\usepackage{epsfig}
\usepackage{float}
\usepackage{multirow}
\usepackage{amsfonts}
\usepackage{amssymb}
\usepackage{amsmath}
\usepackage{tabularx,url,color}
\usepackage{array}
\usepackage[pagebackref=true]{hyperref}
\bibpunct{(}{)}{;}{a}{}{,}

\begin{document}

\title{Scientific performance of on-board analyses for the SVOM X-ray telescope MXT}


\author{F. Robinet \inst{1} \and C. Van Hove \inst{1} \and M. Moita \inst{2} \and S. Crepaldi\inst{3} \and C. Feldman\inst{4} \and A. Fort\inst{3} \and O. Frandon\inst{3} \and D. G\"otz \inst{2} \and P. Maggi \inst{5} \and K. Mercier\inst{3} \and A. Sauvageon \inst{2}}
\institute{
  Universit\'e Paris-Saclay, CNRS/IN2P3, IJCLab, 91405 Orsay, France; {\it florent.robinet@ijclab.in2p3.fr}
  \and
  AIM-CEA/DRF/Irfu/Departement d’Astrophysique, CNRS, Universite Paris-Saclay, Universite Paris Cite, Orme des Merisiers, Bat. 709, Gif-sur-Yvette, 91191, France
  \and
  Centre National d’Etudes Spatiales, Centre spatial de Toulouse, 18 avenue Edouard Belin, 31401 Toulouse Cedex 9, France
  \and
  School of Physics and Astronomy, University of Leicester, Leicester LE1 7RH, UK
  \and
  Observatoire Astronomique de Strasbourg, Universit\'e de Strasbourg, CNRS, 11 rue de l’Universit\'e, 67000 Strasbourg, France
}


\abstract{
  The Microchannel X-ray Telescope on board the Space-based multi-band astronomical Variable Objects Monitor (SVOM) satellite detects and localizes the X-ray afterglow of gamma-ray bursts. One year after the launch, this paper presents the in-flight performance of the scientific analyses conducted by the on-board computer. After summarizing the analysis steps, the paper reviews the on-board results obtained with 15 gamma-ray burst afterglows detected by the telescope between October 2024 and August 2025. For all bursts, the localization uncertainty is estimated to be below 2 arcmin, as required by the mission design. On average, the measured position is found to be 40 arcsec away from the position measured by other experiments with a better sky resolution. Moreover, we show that the on-board analysis provides a precise sky location for the burst only a few seconds after the beginning of the observation. Taking advantage of an efficient very-high-frequency antenna network, this information is quickly collected on the ground and disseminated to other observation facilities. This low-latency strategy is critical for the multi-wavelength and multi-instrument follow-up program of SVOM.
  \keywords{software: data analysis --- gamma-ray bursts --- X-rays: general -- X-rays: bursts}
}
    
\maketitle

\section{Introduction} \label{sec:introduction}

The Microchannel X-ray Telescope (MXT) is described in~\cite{diego}. In section~\ref{sec:methods}, we summarize the on-board analysis methods developed to process the MXT camera images~\cite{aline} and to extract scientific results. For more details, we refer the reader to~\cite{Hussein:2022gwi}. These methods include the photon reconstruction, the source localization, and the estimation of the signal and background counts. The performance is illustrated with a bright and a faint source of X-ray photons (Vela X-1 and S5 0716+714) observed in January 2025. In section~\ref{sec:performance}, the MXT software performance is characterized using gamma-ray burst afterglow data collected by the MXT between June 2024 and August 2025. In particular, the MXT localization performance is estimated and we show that it is consistent with the mission design requirements.

\section{Analysis methods}\label{sec:methods}

\subsection{Photon reconstruction}\label{sec:methods:photons}
An MXT image is thresholded by the camera front-end electronics to only retain the most significant pixels, which are transmitted to the on-board software. The thresholds are tuned pixel-by-pixel using ``dark'' camera images recorded when the telescope wheel is set to a CLOSE position. In Observing Mode, a photon is identified as a cluster of four or fewer contiguous pixels matching a pre-defined list of patterns (see figure 1 in~\cite{Hussein:2022gwi}). A photon is given a position in the focal plane and an energy. The photon energy is obtained by integrating the energies over all the pixels in the cluster. The photon position is estimated using the energy-weighted mean of pixel central positions.

\begin{figure}
  \centering
  \includegraphics[width=0.48\textwidth, angle=0]{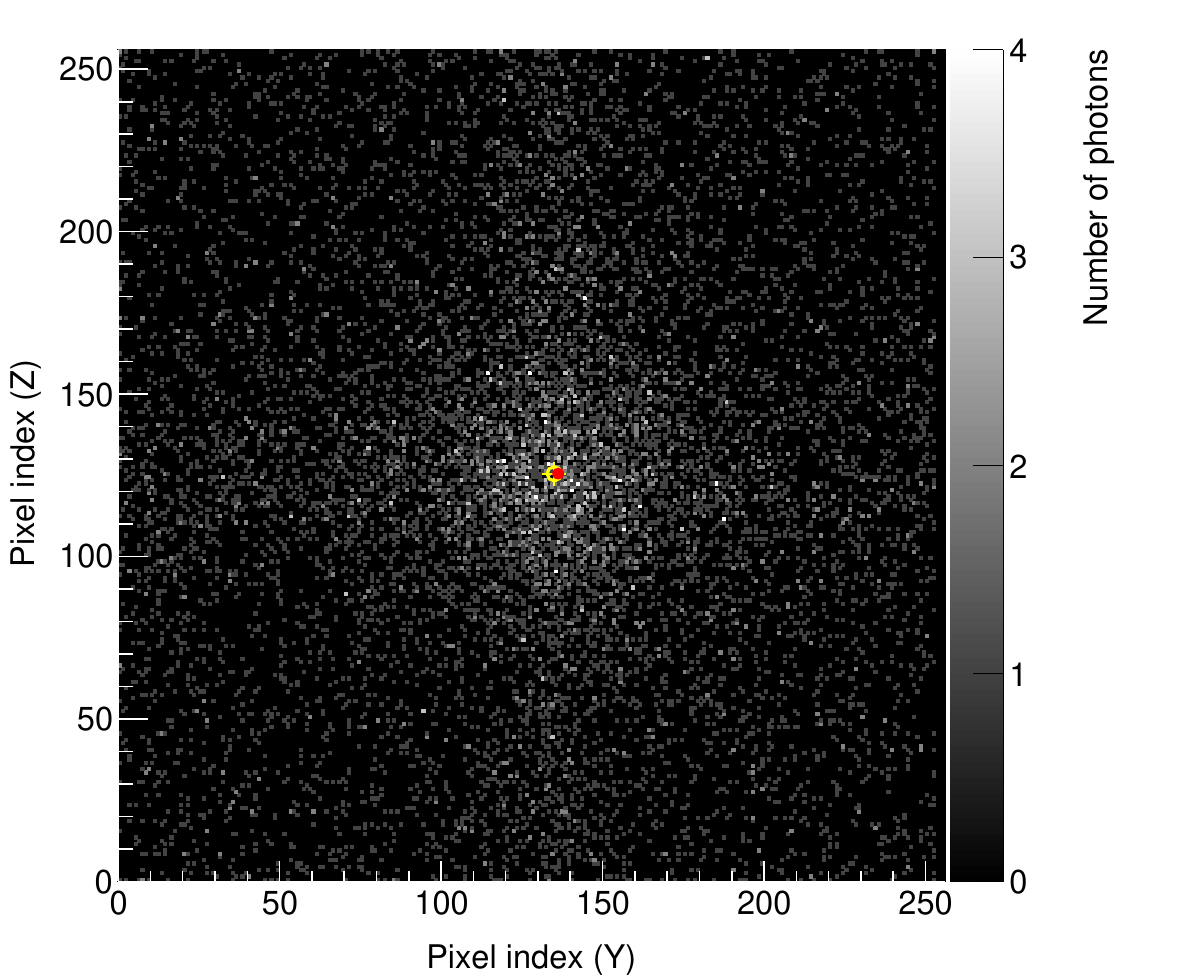}
  \includegraphics[width=0.48\textwidth, angle=0]{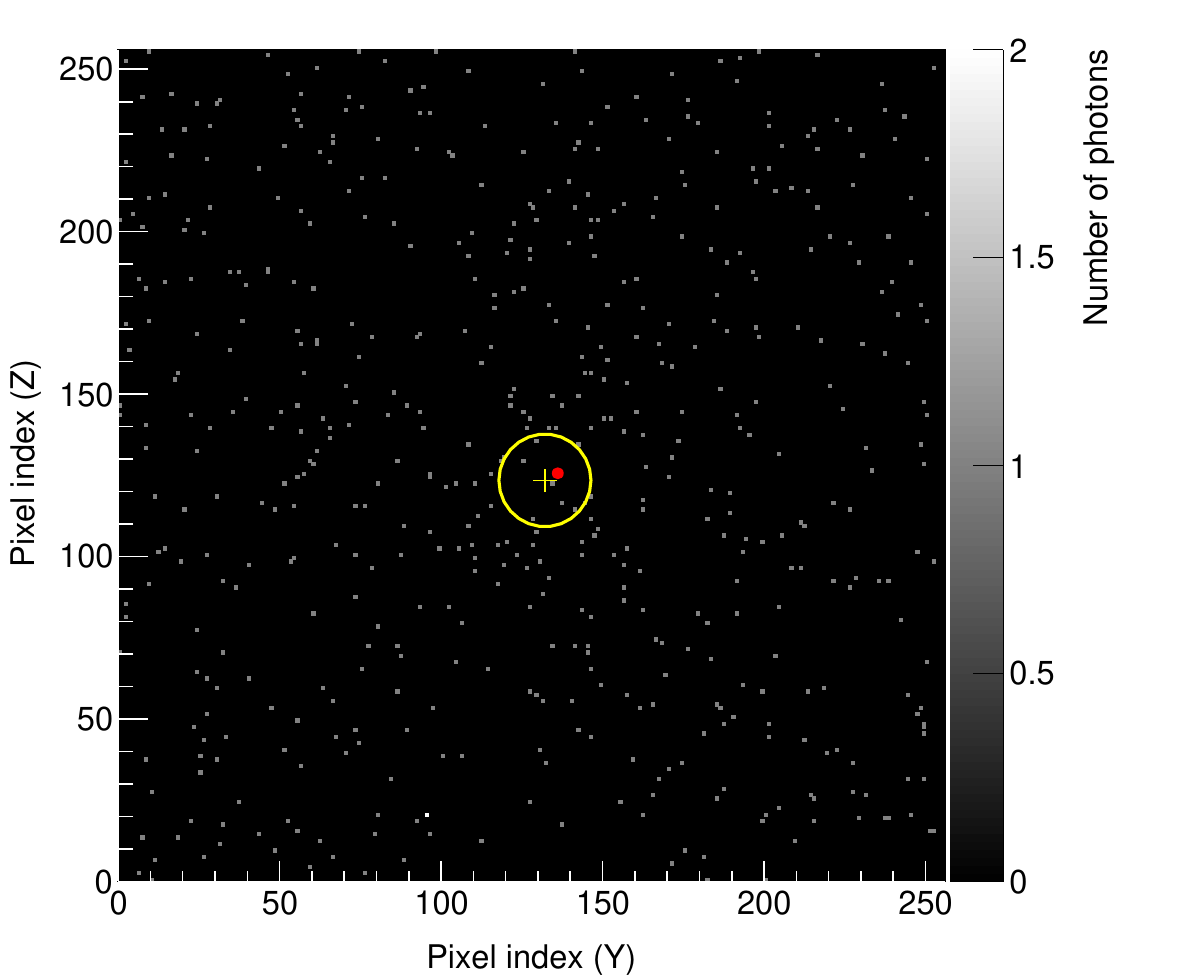}
  \caption{Photon cumulative maps for a bright source (Vela X-1 high-mass X-ray binary, Jan. 12, 2025) in the top panel and a weak source (S5 0716+714 blazar, Jan. 30, 2025) in the bottom panel. The reconstructed source position is represented with a yellow cross and its $R90$ uncertainty with a full yellow circle. The true source position is indicated by a red point.}
  \label{fig:cum_map}
\end{figure}

Reconstructed photons are accumulated onto a $128\times 128$ grid, called the photon cumulative map. The photon cumulative map is updated after each camera image is processed by the on-board software, i.e. every 100~ms. Figure~\ref{fig:cum_map} presents photon cumulative maps obtained when observing a bright (Vela X-1 high-mass X-ray binary, Jan. 12, 2025) and a weak (S5 0716+714 blazar, Jan. 30, 2025) X-ray source. The maps contain 11050 and 544 reconstructed photons respectively.

The list of reconstructed photons is sent within 30 seconds to the ground in a sequence of packets of 35 photons. The first 10 packets are analyzed immediately after reception with a ground analysis pipeline~\cite{pierre} to provide a source position complementing the on-board results.

\subsection{X-ray source localization}\label{sec:methods:localization}
The localization analysis runs continuously on board as a background task. When a localization analysis completes, a new analysis cycle starts immediately. The typical duration for one localization analysis cycle is $\sim 2$~s. The algorithm cross-correlates the current photon cumulative map with the telescope point spread function~\cite{charly}, which describes the expected photon distribution on the focal plane. The cross-correlation analysis takes advantage of the cross-arm shaped point spread function of the MXT to enhance the precision on the direction of incoming photons. Moreover, it integrates the contribution of all X-ray photons in the focal plane, maximizing the sensitivity to faint sources. It is performed in the Fourier domain to optimize the computing time. The resulting cross-correlation map is scanned with a $50\times 50$ pixel-wide window and the cross-correlation peak is identified. This maximum is associated to the source position in the focal plane. The position is converted into a sky position in the MXT reference frame, which is sent to the ground via very-high-frequency channels. On the ground, the position is corrected for alignment biases and the satellite attitude is used to compute the source right ascension and declination in the Earth reference frame (J2000 celestial coordinates). As new photons are collected in the photon cumulative map, the source position is updated and sent every 30 seconds during the first hour, then every 5 minutes for the rest of the observation.

The uncertainty on the localization is measured with the $R90$ quantity, derived from the $90^\mathrm{th}$ percentile of angular differences between measured and true positions. It estimates the position angular uncertainty at a 90\% confidence level. It is derived on board from the source and background counts: see section~\ref{sec:methods:counts}.

Figure~\ref{fig:results} illustrates the localization analysis results for a bright and a faint X-ray source. The sources are observed for several hours with periodic interruptions when the field of view is occulted by the Earth. The localization accuracy, measured with $R90$ (bottom plots), is maximal at the end of the observation for the bright source: it culminates at $R90\sim 30$~arcsec. At this moment, the X-ray source is localized 16.6~arcsec away from the true position, as represented in figure~\ref{fig:cum_map}. For the weak source, the observation is dominated by the background. The localization accuracy oscillates around 200~arcsec. The localization results displayed in the bottom panel of figure~\ref{fig:cum_map} are obtained after 10 minutes of observation when the total number of background photons is four times larger than the number of photons from the source. The source position is found to be 62.0~arcsec away from its true position, which is well within the uncertainty. These two examples show the ability of the cross-correlation analysis to localize accurately an X-ray source in both a signal-dominated and a background-dominated regime.

\subsection{Signal and background reconstruction}\label{sec:methods:counts}

\begin{figure*}
  \centering
  \includegraphics[width=0.48\textwidth, angle=0]{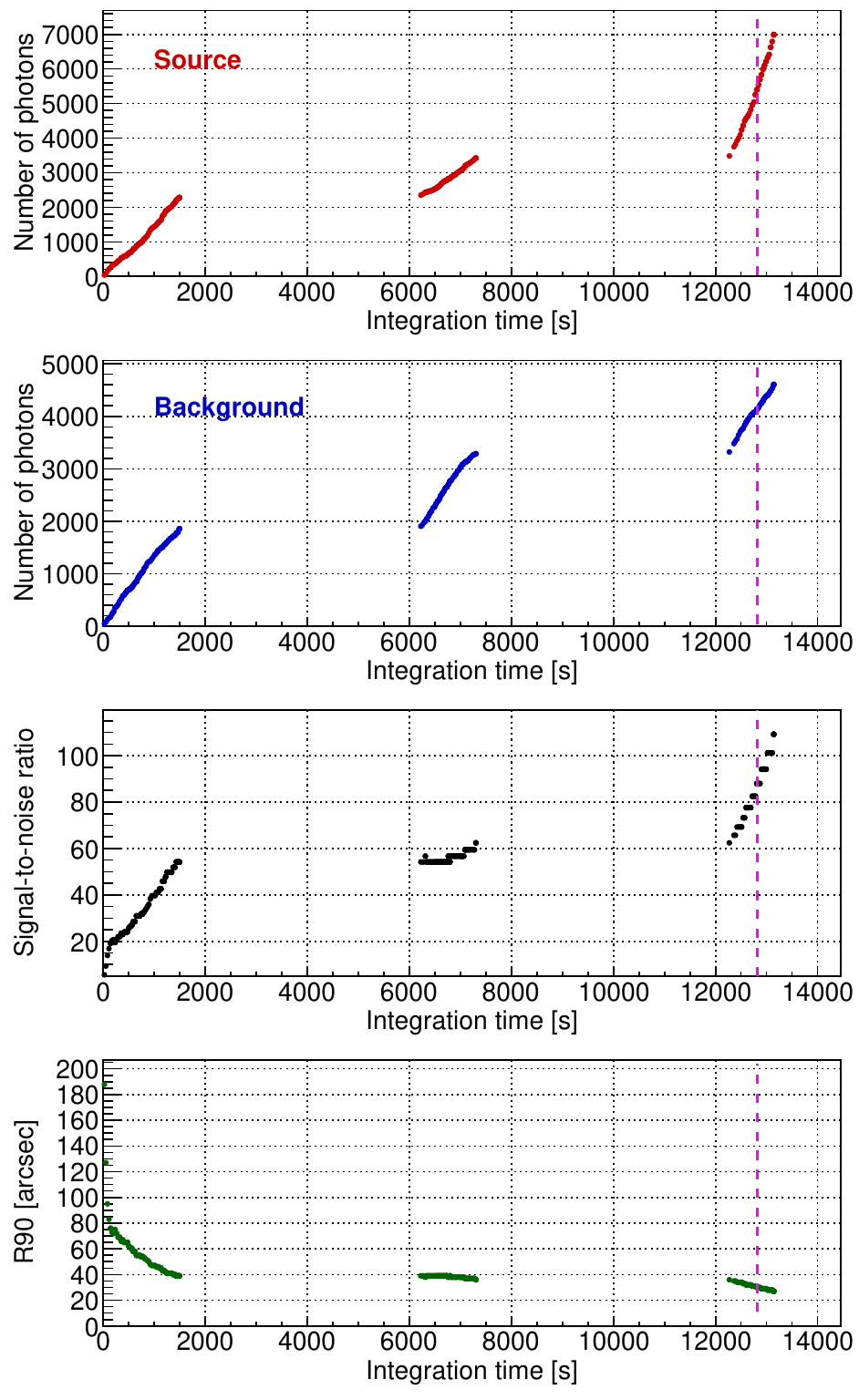}
  \includegraphics[width=0.48\textwidth, angle=0]{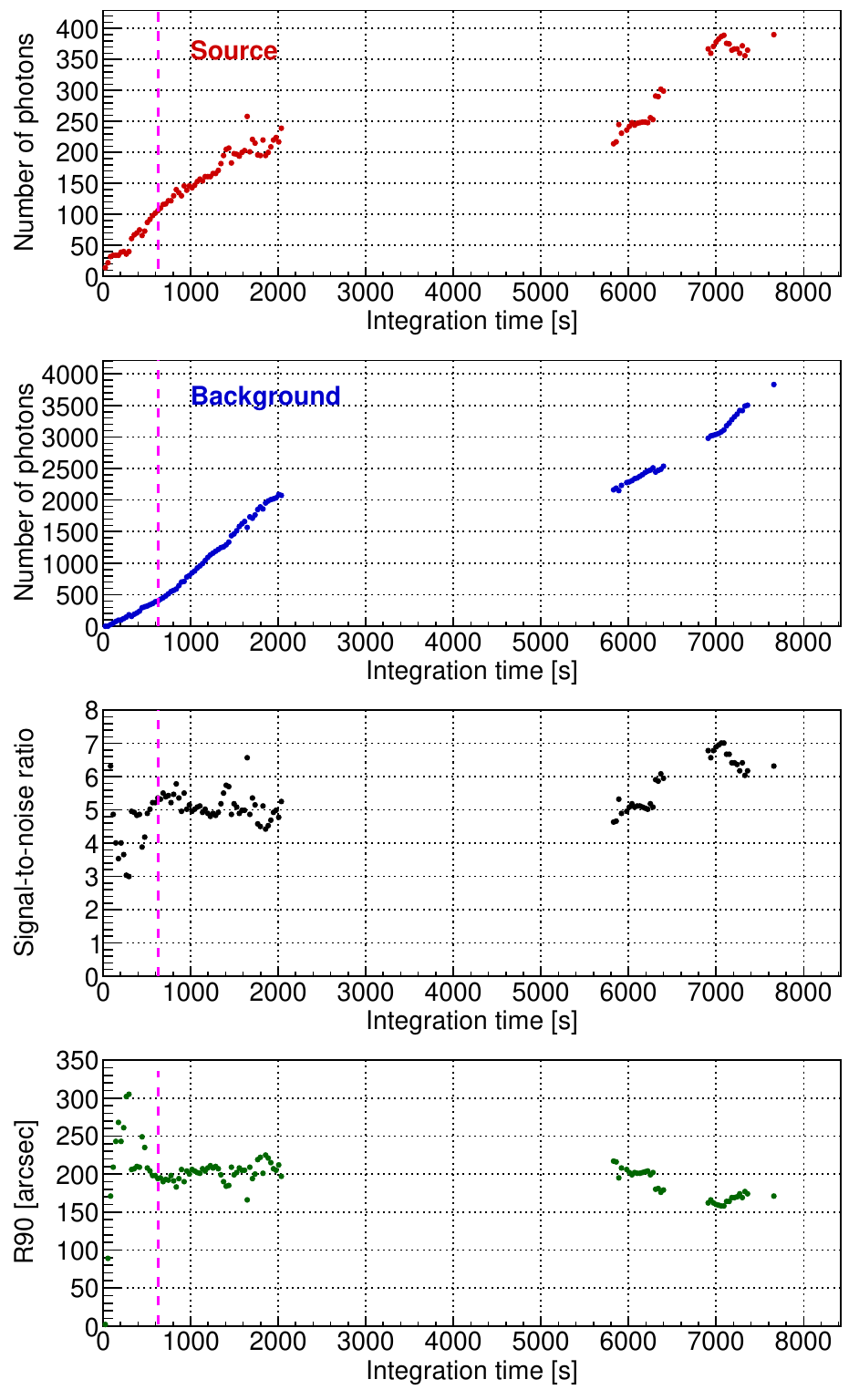}
  \caption{From top to bottom, the cumulative source counts (red), the cumulative background counts (blue), the signal-to-noise ratio (black) and the localization uncertainty $R90$ (green) are plotted as a function of the integration time for a bright source (Vela X-1 high-mass X-ray binary, Jan. 12, 2025) in the left-hand panel and a weak source (S5 0716+714 blazar, Jan. 30, 2025) in the right-hand panel. The vertical magenta dashed lines indicate when the photon cumulative maps in figure~\ref{fig:cum_map} were extracted.}
  \label{fig:results}
\end{figure*}

Due to the peculiar shape of the telescope point spread function, photons collected by the MXT illuminate the entire camera plane. Therefore, it is impossible to isolate a signal-free region to estimate the background counts. An original method has been implemented for the MXT on-board software to separate the X-ray source counts from the background contribution. It relies on the assumption that the background contribution is uniform on the camera plane. Then, the method described in~\cite{Hussein:2022gwi} shows that it is possible to measure the signal and background counts directly from the cross-correlation map constructed to localize the source. This method is all the more effective when the source is well localized.

The signal and background reconstruction is illustrated in the two top panels of figure~\ref{fig:results}. For both the weak and the bright source, the background count rate is estimated to be around 1 photon per second, which is the typical rate for the MXT cosmic X-ray background. This rate is found to be roughly constant over the entire duration of the observation. The bright source (Vela X-1) count rate is not constant, it varies between 0.5 and 5 photons per second. This is compatible with the variability of the X-ray binary. For the weak source, the rate is roughly constant around 0.12 photons per second. It is clear that, in a background-dominated regime, the count reconstruction is less robust, as shown by the fluctuations seen in figure~\ref{fig:results}.

The signal and background counts are used to derive the signal-to-noise ratio, also displayed in figure~\ref{fig:results} (black curve). The localization accuracy $R90$ is analytically computed as an exponential decay function of the signal-to-noise ratio. The analytical expression for $R90$ is obtained from simulated data, as explained in~\cite{Hussein:2022gwi}.

\subsection{Post-launch software patches}\label{sec:methods:patches}

At the time of writing this paper, four patches have been uploaded to the satellite to update the MXT software. They were all developed to mitigate stray-light issues. When the satellite enters or exits the shadow of the Earth, a high rate of photons reflected by the atmosphere can damage the camera. Ground measurements of the impact of stray light on the MXT camera were not fully consistent with the in-orbit behavior~\cite{aline}. As a result, several corrective actions had to be taken. First, protection angles are implemented, under which the camera is switched off. These angles had to be set to high values, hence reducing the observation time drastically. It was realized that the stray-light effect is only constraining when the satellite flies over the day side of the Earth. A first patch, uploaded in September 2024, was developed to track the orbital status of SVOM and to relax the stray-light protection angles on the night side of the Earth. A second patch was uploaded in June 2025 to fine-tune the protection angles: the straylight condition can be further relaxed by checking the elevation angle between the line of sight and the orbital plane.

As explained in sections~\ref{sec:methods:localization} and~\ref{sec:methods:counts}, the on-board analysis assumes the background to be uniform in the camera plane to localize the source and to estimate its intensity. In stray-light conditions, this is no longer true: a background excess is localized near the edges of the detector. If the X-ray source is faint, or if there is no source, this excess can be mis-interpreted as a fake source localized near the edge of the field of view. Since February 2025, the on-board software has been protected against this effect by rejecting sources localized far away from the camera center. It is important to note that MXT observations of gamma-ray bursts are triggered by ECLAIRs, the localization accuracy of which guarantees that the real source is always positioned near the center of the focal plane.

Finally, in stray-light conditions, the input photon flux increases significantly. It was realized that the clustering algorithm presented in section~\ref{sec:methods:photons} took too much time to complete, triggering a software fail-safe mode. This was fixed on February 2025 with an optimized clustering algorithm.

\section{Scientific performance} \label{sec:performance}

\subsection{Observation time} \label{sec:performance:obstime}

The MXT operation is organized around working modes which are automatically managed by the on-board software. The observation mode, i.e. when scientific data are collected, represents $\sim 35\%$ of the MXT operation time. For a large fraction of the time ($\sim 45\%$), the MXT is occulted by the Earth or protected when in stray-light conditions. Moreover, the MXT is switched off when crossing the South Atlantic Anomaly ($\sim 15\%$). The remaining $5\%$ of the time is reserved for technical operations.

As presented in section~\ref{sec:methods:patches}, several software patches have been uploaded to improve the observation time of the MXT before and after an Earth occultation. After patching the software to relax the stray-light protection, the time when the MXT records scientific data has improved by $\sim 6\%$. This number is subject to large uncertainties because the last patch was uploaded recently (June 2025).

\subsection{Gamma-ray bursts} \label{sec:performance:grb}

From the SVOM launch until August 15, 2025, 16 gamma-ray burst afterglows were detected by the MXT~\cite{GRB_241018A,GRB_241018A_MXT,GRB_241212A,GRB_241217A,GRB_250108A,GRB_250205A,GRB_250317B,GRB_250327B,GRB_250402A,GRB_250506A,GRB_250507A,GRB_250512B,GRB_250704A,GRB_250706B,GRB_250713A,GRB_250806A,GRB_250813B}. All but one were confidently detected on board. An X-ray transient is said to be detected with confidence by the MXT if the signal-to-noise ratio is larger than 4 and if the estimated number of photons from the source (see section~\ref{sec:methods:counts}) is larger than 100.

\begin{figure}
  \centering
  \includegraphics[width=0.5\textwidth, angle=0]{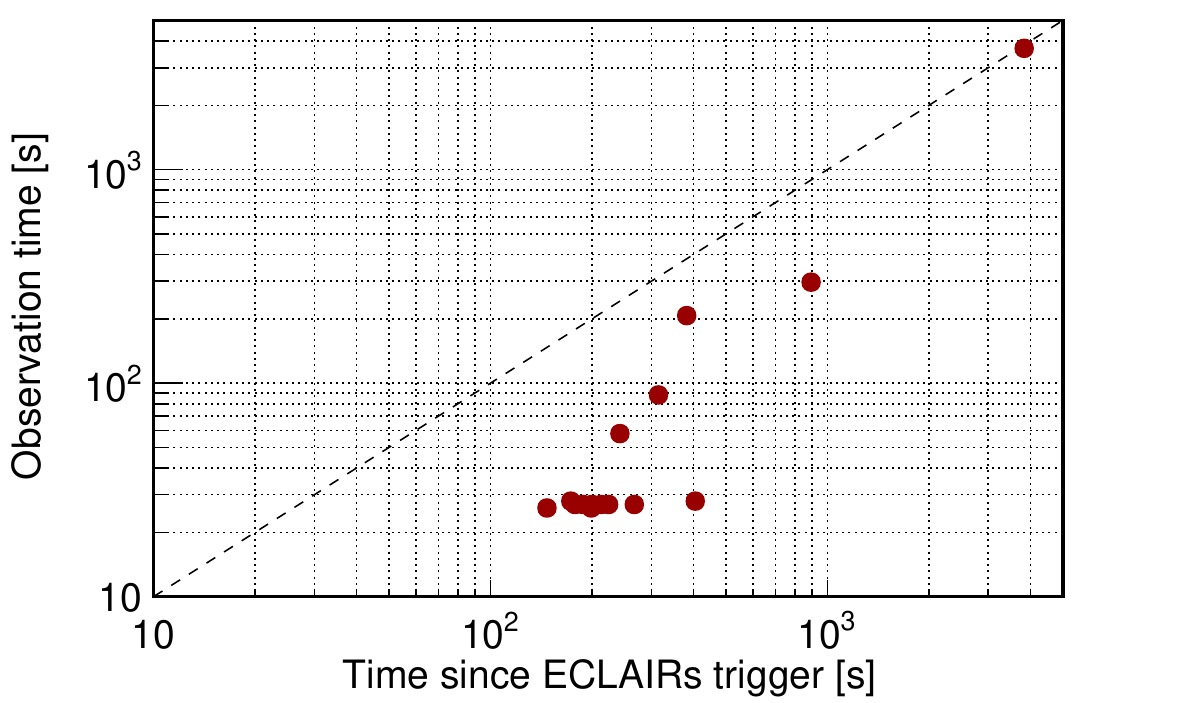}
  \caption{Observation time needed to confidently detect 15 gamma-ray burst afterglows with the MXT. It is plotted as a function of the time since the original ECLAIRs trigger.}
  \label{fig:grb_timedelay}
\end{figure}
To coordinate the multi-wavelength and multi-instrument follow-up of a gamma-ray burst and its afterglow, it is critical to report and publish the precise sky localization of the source rapidly. In figure~\ref{fig:grb_timedelay}, we show how much observation time was needed to detect with confidence 15 gamma-ray burst afterglows with the MXT. The observation time starts when the telescope wheel moves to the OPEN position after the satellite slew and when the observation is possible. It is plotted against the time since the original ECLAIRs trigger. For 10 gamma-ray bursts, the observation time is below 30~s: this means that the localization results were received in the first very-high-frequency packet sent to the ground (see section~\ref{sec:methods:localization}). The time since ECLAIRs trigger is always larger than the observation time because it also includes the satellite slew, which can last up to several minutes. Moreover, the start of observation can be delayed if the slew ends while the MXT is occulted by the Earth. In figure~\ref{fig:grb_timedelay}, the detection can be delayed significantly: GRB 241212A was detected after one full hour. In this case, the observation was interrupted after only having recorded a few seconds of data, due to the crossing of the South Atlantlic Anomaly followed by an Earth occultation. Even if localization results were available on board, they could not be transmitted to the ground. When the MXT was back online, one hour later, a new localization cycle had ended and results had been tagged with a very long observation time~\footnote{Luckily, for GRB 241212A, the General Coordinates Network (GCN) circular was drafted with localization results obtained with the ground pipeline~\cite{pierre} using the list of photons received before the MXT went dark.}.

\begin{figure*}
  \centering
  \includegraphics[width=0.95\textwidth, angle=0]{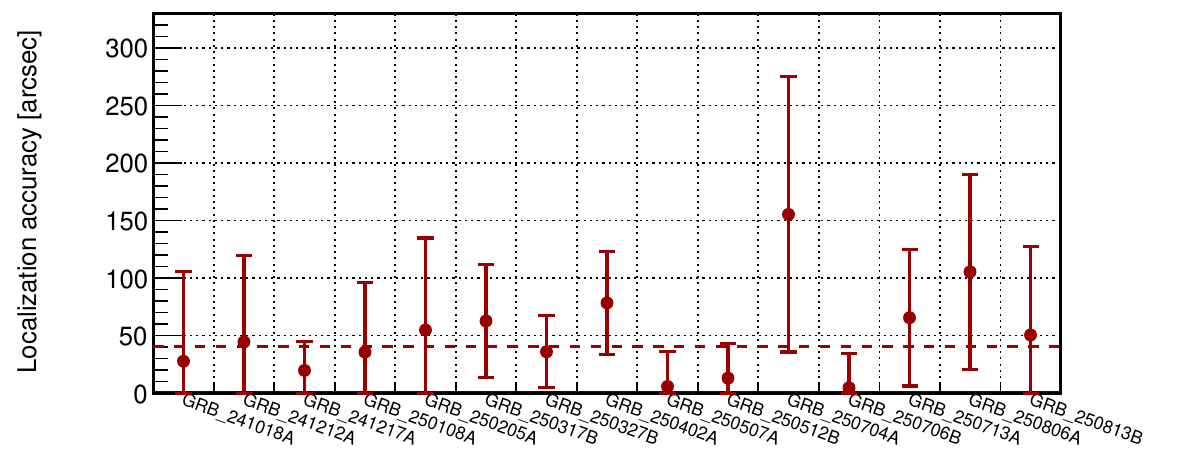}
  \caption{Localization accuracy associated with the 15 first gamma-ray burst afterglows detected by the MXT. The points indicate the angular difference between the position measured on board and the position measured by other instruments (Swift/XRT~\cite{Gehrels:2004qma}, SVOM/VT~\cite{Wang:2009pn}, EP/FXT~\cite{Yuan:2022fpj,Zhang:2025vgv}) known to localize a source with a precision of a few arcsec. The error bars are computed as the quadratic sum of uncertainties at a 90\% confidence level, including systematics. The horizontal dashed line (40.3~arcsec) marks the median value of angular difference for all gamma-ray bursts.}
  \label{fig:grb_locaccuracy_grb}
\end{figure*}
In figure~\ref{fig:grb_locaccuracy_grb}, the on-board localization accuracy is characterized. For each gamma-ray burst, we report on the angular difference between the on-board position minimizing $R90$ and the best position reported by other instruments. The two positions are consistent if the point in figure~\ref{fig:grb_locaccuracy_grb} is compatible with 0. The on-board localization uncertainty is below 2~arcmin for all gamma-ray bursts, which was a design requirement for the MXT. For four gamma-ray bursts, it is negligible ($<2$~arcsec) compared to the systematic uncertainty, which is about 25~arcsec. Considering all the gamma-ray bursts, the median value for the angular difference is 40.3~arcsec.

\subsection{GRB 241217A: a bright gamma-ray burst} \label{sec:performance:grbex}

\begin{figure*}
  \centering
  \includegraphics[width=0.99\textwidth, angle=0]{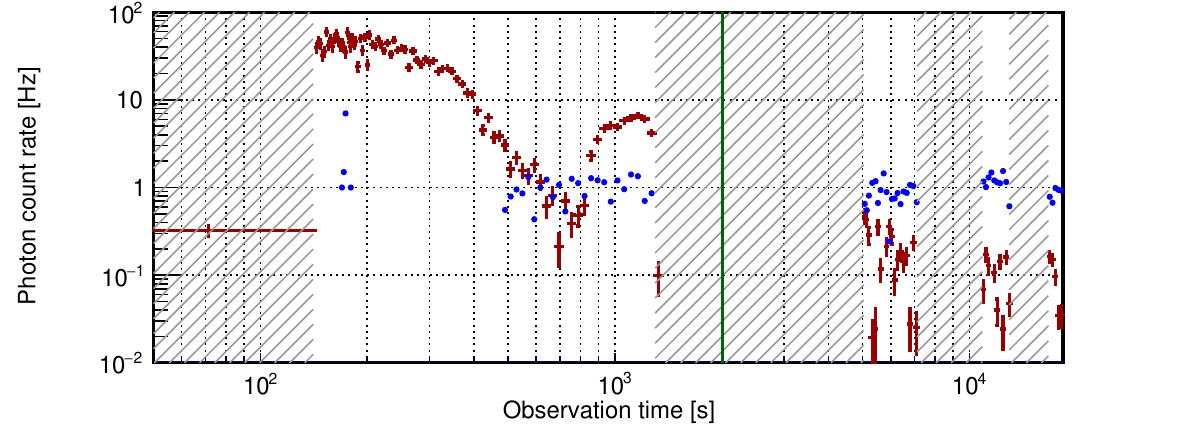}
  \caption{MXT observation of GRB 241217A. The background-subtracted light curve, as calculated on board, is plotted as a function of time (red points). The corresponding background contribution is indicated with the blue points. The hatched areas mark the observation interruptions. During the second interruption, the SVOM satellite was slewed a second time (green vertical line) to center the source in the field of view.}
  \label{fig:grb241217a_lc}
\end{figure*}

On December 17, 2024, a bright gamma-ray burst was detected by the ECLAIRs detector. After 150~s of slew, the afterglow was followed by the MXT over 4 orbits, accumulating more than 5 hours of data. In the first observation period, the signal flux culminates at a rate of 60 photons per second, as shown in figure~\ref{fig:grb241217a_lc}. The signal-to-noise ratio reaches out-of-range values ($\gg 100$). In this regime, the localization accuracy saturates and is only limited by systematic uncertainties. The signal and background counts are reconstructed with precision. The afterglow light curve, after an initial decay, exhibits a flux increase between 700~s and 1300~s. During the first Earth occultation period, the satellite was slewed a second time based on the MXT localization results, to center the source in the MXT and VT field of views. After the second slew, the MXT localization analysis is reset, meaning that all internal counters are set back to 0. A new observation starts after 5000~s. The afterglow is still detectable by the MXT, as witnessed by the decreasing photon rate in figure~\ref{fig:grb241217a_lc}. In the third and fourth orbits, the X-ray source is no longer detectable: the source count rate estimated on board saturates at around $\sim 0.1$~photons per second, which is the effective detectability limit for the MXT.

\section{Conclusion} \label{sec:conclusion}

In this paper, we characterize the scientific performance of the on-board software of the MXT over one year of observation. Multiple gamma-ray bursts and other sources have been followed up with the telescope. The on-board software is able to detect astrophysical X-ray photons down to a rate of 0.1~photons per second, allowing for the MXT to be sensitive to faint sources. The X-ray source is promptly localized on board as the result of a cross-correlation analysis. All the gamma-ray burst afterglows detected so far by the MXT have been localized with an accuracy better than 2 arcmin, as required by the mission design. On average, the sky position estimated by the MXT is 40.3 arcsec away from the true source position. The localization results are sent to the ground via a very-high-frequency network. For 66\% of the gamma-ray bursts, a sky position is received within 30~s.

After the SVOM launch, the MXT on-board software has been patched four times to mitigate stray-light issues. The main outcome is an increased observation time for the telescope after the patches are active. After one year, the MXT operates nominally and the on-board software has reached a stable and mature state.

\begin{acknowledgements}
  The Space-based multi-band astronomical Variable Objects Monitor (SVOM) is a joint Chinese-French mission led by the Chinese National Space Administration (CNSA), the French Space Agency (CNES), and the Chinese Academy of Sciences (CAS). We gratefully acknowledge the unwavering support of NSSC, IAMCAS, XIOPM, NAOC, IHEP, CNES, CEA, and CNRS. We acknowledge the strong involvement of the CNES team during the commissioning phase: Karine Mercier, Adrien Fort, Olivier Frandon, and Stefano Crepaldi.
\end{acknowledgements}

\bibliographystyle{raa}
\bibliography{references}

\end{document}